# Differentiating Eulerian and Lagrangian Tendencies in the Ocean Interior via a Dynamical Overturning Decomposition

Lei Han [a, b]

[a] *China-ASEAN College of Marine Sciences, Xiamen University Malaysia, Sepang, Malaysia*

[b] *College of Ocean and Earth Sciences, Xiamen University, Xiamen, China*

Corresponding author: Lei Han (lei.han@xmu.edu.my)

**Abstract**

Repeat hydrographic observations provide essential constraints on long-term changes in the ocean interior, such as warming and cooling trends. However, attributing these Eulerian signals to either reversible isopycnal heaving or irreversible diabatic transformations remains a fundamental challenge. This ambiguity arises not only from limited velocity observations, but also from the lack of a diagnostic framework capable of explicitly disentangling Lagrangian transformation from Eulerian variability. Recognizing that Eulerian and Lagrangian tendencies can be represented through distinct perspectives of the overturning circulations, we apply a dynamical overturning decomposition to state-of-the-art reanalysis products in the Atlantic, Indo-Pacific, and South China Sea to investigate their long-term behavior. The utility of this framework is supported by strong correlations between independently derived kinematic and thermodynamic indices, indicating a tight coupling between advective and isopycnal-heaving transports. Our analysis reveals distinct dynamical regimes in which similar Eulerian tendencies arise from fundamentally different Lagrangian processes. In particular, we identify (i) previously undiagnosed diapycnal downwelling, despite its prediction by bottom-intensified turbulent dissipation, (ii) coexistence of apparent Eulerian upwelling and diapycnal downwelling, (iii) an Atlantic sub-overturning cell linked to intermediate water-mass formation, and (iv) cases where Eulerian densification masks Lagrangian lightening. These results demonstrate that Eulerian trends alone can be misleading indicators of reversible and irreversible behaviors. By explicitly separating adiabatic heaving from diabatic processes, this framework offers a physically consistent diagnostic for interpreting interior ocean changes and establishes a dynamical basis for assessing the reversibility of such changes. This distinction is critical for understanding long-term ocean variability and its role in climate, particularly in evaluating ocean heat uptake and storage.

**Significance statement**

Understanding how the ocean is changing is essential for predicting future climate. However, interpreting ocean observations is not always straightforward. When measurements show changes at a given depth, it is often unclear whether the water itself is changing—such as warming or cooling—or whether it is simply being displaced by ocean motion. In this study, we develop a new approach to distinguish between these two possibilities. By applying this method to multiple ocean regions, we find that the same observed change can arise from very different underlying processes. In some cases, water can appear to move upward at a fixed location while it is actually becoming denser and moving downward. This seemingly contradictory behavior results from large-scale vertical shifts of ocean layers. These findings highlight that observed changes in the ocean can be misleading if such vertical movements are not properly accounted for. Our work provides a more reliable way to interpret ocean observations and offers new insight into how the ocean stores heat and responds to climate change.

## 1. Introduction

Emerging evidence has begun to overturn the notion that the deep ocean is quiescent. Repeated hydrographic sections in the Pacific have revealed consistent deep-ocean changes (Sloyan et al. 2013), and volume budgets below deep, cold isopycnals are reportedly "not in steady state" (Purkey and Johnson 2012). Comparisons between the 1870s *HMS Challenger* data and modern measurements suggest a centennial cooling of ~0.02°C in the deep Pacific and a warming tendency in the deep Atlantic (Gebbie and Huybers 2019). More recently, global deep and abyssal temperature trends have been mapped using Deep Argo and CTD measurements (Johnson and Purkey 2024), alongside documentation of basin-scale trends in overall isopycnal volumes identified in ocean reanalyses (Monkman and Jansen 2024).

Concurrently, vertical profiles of turbulent fluxes offer complementary constraints on local density tendencies. Field measurements since the 1990s have discovered significantly elevated diffusivities near the seafloor, particularly in areas with distinct topographic features and regions of heightened internal-wave energy (e.g., Toole et al. 1994; Kunze et al. 2006; Lee et al. 2006; Waterman et al. 2013; Waterhouse et al. 2014; Voet et al. 2015; Mashayek et al. 2017; Kunze 2017a, b). Crucially, a bottom intensification of turbulent dissipation rates or vertical diffusive fluxes has been consistently observed across various field campaigns (e.g., Polzin et al. 1996; Morris et al. 2001; St. Laurent et al. 2001; Waterman et al. 2013; Mashayek et al. 2017). Physically, the convergence associated with these vertical diffusive flux gradients drives a local tendency toward densification within the water column.

This poses a fundamental question regarding the dynamical nature of these observed density tendencies. Do these signals reflect reversible isopycnal heaving, or do they represent irreversible water-mass transformation? While fixed-point observations effectively capture the Eulerian tendency of the ocean environment, they are inherently unable to disentangle the specific physical origins of these local changes. As a result, it remains unclear whether observed Eulerian trends provide a faithful measure of irreversible behavior, or whether they may instead arise from largely reversible adiabatic processes.

To address this challenge, we employ a recently developed dynamical decomposition framework designed for the Meridional Overturning Circulation (MOC). This framework integrates three definitions of streamfunction: the advective (Eulerian) MOC, the sloshing MOC, and the diapycnal MOC streamfunctions. While the first two capture the fixed-coordinate volume transport and the isopycnal inflation/deflation rate, respectively, their residual—the diapycnal or diffusive component—explicitly isolates cross-isopycnal transport, thereby quantifying the diabatic contribution to the overturning system. This formulation provides a natural dynamical separation between Eulerian variability and Lagrangian transformation.

This framework has demonstrated its efficacy in providing new insights to the notable seasonality of the Indian Ocean MOC (Han 2021) and in characterizing the variability and meridional connectivity of the Atlantic MOC (AMOC; Han 2023a, b, 2025). Notably, by identifying the sloshing nature of AMOC at short timescales, Han (2023a) suggested that the system is inherently more resilient than previously inferred. Although this perspective diverged from the then-prevailing narrative of a secular AMOC decline (e.g., Rahmstorf et al. 2015; Caesar et al. 2018; Boers 2021; Caesar et al. 2021), it has since been substantiated by crucial observational updates. Volkov et al. (2024) corrected a previously reported spurious weakening trend in the continuous monitoring array, revealing a significantly more stable AMOC that aligns with the resilience predicted by our dynamical framework.

Physically, a downwelling sloshing MOC corresponds to a downward shift of isopycnals, manifesting as Eulerian lightening (or Eulerian cooling if tracking isotherms) at fixed locations. By isolating the sloshing component from the total advective overturning, we resolve the diapycnal circulation, which represents irreversible water-mass transformation along parcel trajectories. Consequently, an upwelling diapycnal MOC thus signifies Lagrangian lightening driven by cross-isopycnal volume flux. The relative contributions of these components therefore provide a direct dynamical measure of the reversibility of Eulerian variability.

Implementation of this framework necessitates velocity fields alongside hydrographic data, currently limiting its application to model-based products. However, with the rigorous assimilation of in-situ data, modern reanalysis datasets exhibit increasing fidelity to the real ocean. We therefore utilize two state-of-the-art reanalysis products as testbeds to demonstrate the framework's utility across diverse regimes, including two major ocean basins (the Atlantic and Indo-Pacific Oceans) and a marginal sea (the South China Sea). By extending this framework from variability to long-term tendencies, we identify distinct dynamical regimes in which similar Eulerian signals arise from fundamentally different Lagrangian processes. These regimes include cases where apparent upwelling coincides with diapycnal downwelling, as well as situations in which Eulerian signals mask underlying transformations. This result underscores the imperative of jointly examining both perspectives when assessing changes in the deep-ocean state.

The remainder of the paper is structured as follows. Section 2 details the datasets and the sloshing-diapycnal decomposition methodology. Section 3 presents the findings from applying the framework to the three basins. Finally, Section 4 summarizes the findings and discusses their broader implications.

## 2. Data and Methodology

### 2.1. Data

#### 2.1.1. ECCO: coarse-resolution global ocean state estimate

We adopt the Estimating the Circulation and Climate of the Ocean version 4 release 3 (ECCO v4r3) (Forget et al. 2015; Fukumori et al. 2017). This data assimilation product is generated by optimally constraining the numerical solution of an ocean general circulation model with over 1 billion observations (Rousselet et al. 2020, 2022). The ECCO v4r3 dataset (hereafter referred to as ECCO) spans the 24-year period (1992–2015) with a monthly temporal resolution, a nominal horizontal resolution of 1°, and 50 vertical levels.

In this study, we employ ECCO to diagnose basin-scale Eulerian and Lagrangian tendencies in the Atlantic and Indo-Pacific Oceans. Distinguished for its strict dynamical consistency, the ECCO solution is uniquely capable of supporting closed budget analyses, such as volume and tracer transports. Consequently, it has been extensively utilized to investigate MOC systems in the Atlantic Ocean (e.g., Cabanes et al. 2008; Baehr et al. 2009; Evans et al. 2017; Jackson et al. 2019; Smith and Heimbach 2019; Kostov et al. 2021; Han 2023a, b, 2025), the Indo-Pacific basin (e.g., Rousselet et al. 2021; Han 2021; Rogers et al. 2023), and the global ocean (e.g., Cessi 2019; Monkman and Jansen 2024).

#### 2.1.2. GLORYS12v1: high-resolution reanalysis dataset

GLORYS12V1 (hereafter GLORYS) is an eddy-resolving global ocean reanalysis produced by Mercator Ocean International under the Copernicus Marine Environment Monitoring Service (CMEMS). It is generated using the NEMO ocean–sea-ice model, forced by ERA-Interim atmospheric fields, and provides data from January 1993 to December 2024 (32 years) at a horizontal resolution of 1/12° with 50 vertical levels (Lellouche et al. 2021).

In this study, we utilize monthly averaged GLORYS fields to investigate the South China Sea (SCS), the marginal basin linking the Indian and Pacific Oceans. We employ this high-resolution reanalysis specifically because the confined spatial extent and complex topography of SCS require finer resolution than is typically available in coarse, basin-scale products such as ECCO.

GLORYS has undergone extensive validation at both global and regional scales (Lellouche et al. 2021). In a comparative assessment of eight global ocean reanalyses over continental shelves, GLORYS demonstrated the strongest overall performance across most metrics (Castillo-Trujillo et al. 2023). Of particular relevance to its application in this study, GLORYS shows superior skill relative to other prominent products (e.g., ORAS5 and SODA) in reproducing intermediate and deep circulation patterns within the SCS (Liu et al. 2025), and has been recently applied to studies of the deep SCS circulation (Li et al. 2025). However, in our analysis, outliers are identified in approximately 3% of the 32-year monthly time series when diagnosing isopycnal depths in the SCS, as detailed in Section 3.3.

## 2.2. Methodology

### 2.2.1. Advective and sloshing MOC streamfunctions

The advective MOC streamfunction (denoted as $\psi_{\mathrm{adv}}$) quantifies the zonally-integrated volume flux of water parcels in the meridional or vertical directions. In this study, two forms of advective MOC streamfunctions are employed, depending on the dataset: the Eulerian MOC streamfunction ($\psi_{Eul}$) and the residual MOC streamfunction ($\psi_{res}$).

The Eulerian MOC streamfunction in potential-density space is computed by vertically integrating the zonally-integrated Eulerian meridional velocity ($v$) over all densities greater than a specified value ($\sigma_\theta$),

$$\psi_{Eul}(y, \sigma_\theta, t) = \int_{\infty}^{\sigma_\theta} d\sigma \int_{x_w}^{x_e} v(x, y, \sigma, t) dx. \tag{1}$$

It represents the net meridional volume transport, across a zonal section at latitude $y$, of waters denser than $\sigma_\theta$ advected by the Eulerian flow.

The residual MOC incorporates parameterized eddy transport by including an additional eddy-induced (bolus) velocity ($v^*$) in the integration,

$$\psi_{res}(y, \sigma_\theta, t) = \int_{\infty}^{\sigma_\theta} d\sigma \int_{x_w}^{x_e} (v + v^*) dx. \quad (2)$$

The residual MOC reflects the net overturning transport from the combined Eulerian-mean and eddy-induced circulations (e.g., Marshall and Radko 2003; Nurser and Lee 2004; Ito and Marshall 2008; Munday et al. 2013), and closely resembles the Eulerian MOC over much of the basin interior where isopycnals are relatively flat (Ito and Marshall 2008). Accordingly, the residual streamfunction is more appropriate for coarse-resolution products with parameterized eddy effects (e.g., ECCO), whereas for eddy-resolving products (e.g., GLORYS), the Eulerian streamfunction alone is generally sufficient.

In contrast, the sloshing MOC streamfunction, $\psi_{slo}$, captures the vertical displacement of isopycnals rather than the physical motion of water parcels. It is obtained by integrating the vertical displacement rate of the isopycnal $\sigma_\theta$, $w_{iso}$, either northwards:

$$\psi_{slo}(y, \sigma_\theta, t) = \psi_{slo}(y_S, \sigma_\theta, t) + \int_{y_S}^{y} dy' \int_{x_w}^{x_e} w_{iso}(x, y', \sigma_\theta, t) dx, \quad (3a)$$

or southwards (Han 2021, 2023a):

$$\psi_{slo}(y, \sigma_\theta, t) = \psi_{slo}(y_N, \sigma_\theta, t) - \int_{y}^{y_N} dy' \int_{x_w}^{x_e} w_{iso}(x, y', \sigma_\theta, t) dx. \quad (3b)$$

Here, $w_{iso}$ is derived from the temporal evolution of isopycnal depths. For monthly data, it is approximated as

$$w_{iso}(\sigma_\theta, m) = \frac{1}{T}[h(\sigma_\theta, m) - h(\sigma_\theta, m-1)], \qquad m = 1,2, \ldots, N \quad (4)$$

where $h(\sigma_\theta, m)$ denotes the depth of the isopycnal $\sigma_\theta$ in month $m$, and $T$ is set to 30 days. The density variable $\sigma_\theta$ may correspond to potential densities, $\sigma_0$, $\sigma_2$, or $\sigma_4$, referenced to the surface, 2,000 dbar, and 4,000 dbar, respectively. An upward $\psi_{slo}$ indicates rising isopycnals and corresponds to local Eulerian densification, whereas downward displacement corresponds to Eulerian lightening.

The main methodological uncertainty of our framework arises from the estimation of isopycnal depths, $h(\sigma_\theta, m)$, which relies on linear interpolation of the density field between adjacent vertical grid points. This approach is widely adopted in the community for similar purposes (e.g., Monkman and Jansen 2024), and is not expected to affect the qualitative conclusions.

Concerning the boundary conditions in Eq. (3) (i.e., $\psi_{slo}(y_S)$ and $\psi_{slo}(y_N)$), integration from a solid boundary is straightforward, as the boundary value vanishes. In contrast, integration from an open meridional boundary (e.g., in the Atlantic) requires specifying the advective MOC streamfunction as the boundary condition to ensure consistency with a well-defined diapycnal overturning component. The rationale for this treatment is provided in the derivation of Eq. (6) at the end of Section 2.2.2.

Although the term "sloshing" typically refers to oscillatory, seiche-like behavior—which is appropriate for describing seasonal or interannual variability of the MOC system (Han 2021, 2023a, b, 2025)—in the context of long-term trends, it is used to refer to a monotonic or unidirectional shift of isopycnals over the analysis period. Importantly, reversibility cannot be inferred from the sloshing component alone, as isopycnal displacement can arise from both adiabatic and diabatic processes. A proper assessment of reversibility therefore requires consideration of the diapycnal component, as defined in the next section.

### 2.2.2. Diapycnal MOC streamfunction

The advective MOC streamfunction ($\psi_{Eul}$) does not inherently represent water-mass transformation. As demonstrated in the Indian Ocean, a substantial fraction of the advective MOC can be adiabatic in origin (Han 2021). It is the residual between the advective MOC ($\psi_{Eul}$) and the sloshing MOC ($\psi_{slo}$) that reflects the rate at which water parcels cross density surfaces, i.e., the diapycnal transformation rate. This cross-isopycnal transport is exclusively driven by diffusive or diabatic effects (Han 2025).

To quantify this transport, we define the diapycnal MOC component, $\psi_{dia}$, as follows:

$$\psi_{dia} = \psi_{Eul} - \psi_{slo}, \tag{5}$$

This decomposition has been validated in both the Indian and Atlantic Oceans, where it effectively separates adiabatic and diabatic contributions to short-term MOC variability.

Diapycnal transport has traditionally been diagnosed using a range of approaches based on both model output and observations. In the Eulerian framework, water-mass transformation rates are commonly inferred from thermodynamic budgets (e.g., de Lavergne et al. 2016; Ferrari et al.

2016; Cimoli et al. 2023; Monkman and Jansen 2024). Alternatively, Lagrangian approaches track water-mass pathways and transformations directly (e.g., Shu et al. 2014; Rousselet and Cessi 2022).

The present framework offers a complementary perspective by explicitly accounting for the time evolution of isopycnals, allowing isopycnal heaving to be separated from diapycnal transformation within a purely Eulerian framework, without requiring explicit trajectory calculations. In this study, the decomposition framework is formulated in density space, rather than physical space as in previous studies (Han 2021, 2023a). It is conceptually equivalent to the volume-budget approach of Monkman and Jansen (2024), although the two frameworks differ slightly in their interpretation of diapycnal transport. A detailed discussion of this distinction is provided in Section 3.2.2.

Physically, $\psi_{dia}$ quantifies the transport of water parcels across isopycnals and therefore diagnoses their Lagrangian buoyancy change. A positive $\psi_{dia}$ indicates that water parcels ascend faster (or descend slower) than the isopycnals, implying an upward crossing of density surfaces, and hence a gain in buoyancy (Lagrangian lightening). Conversely, a negative $\psi_{dia}$ corresponds to Lagrangian densification.

The relationship between the advective, sloshing, and diapycnal transports can be understood through a volume budget framework, as schematized in Figure 1. Consider a control volume bounded by a specified isopycnal surface ($\sigma_\theta$), solid boundaries (the seafloor and eastern and western coasts), and two meridional sections ($y_0$ and $y_1$). The volume tendency arises solely from the vertical movement of its only deformable boundary—the upper isopycnal surface—and can be expressed as a heaving transport ($T_{hv}$), obtained by integrating the isopycnal displacement rate ($w_{iso}$) over the horizontal area. The term $T_{hv}$ is equivalent to the *dV/dt* term defined by Monkman and Jansen (2024).

This volume tendency is balanced by fluxes through the open boundaries, including meridional transport ($\psi_{\mathrm{adv}}$, either $\psi_{Eul}$ or $\psi_{res}$) across the zonal sections and diapycnal transport ($\psi_{dia}$) across the upper surface due to diabatic processes. The volume balance can therefore be written as

$$T_{hv} = \psi_{\mathrm{adv}}(y_1, \sigma_\theta) - \psi_{\mathrm{adv}}(y_0, \sigma_\theta) - \psi_{dia}\,, \qquad (6a)$$

and thus

$$\psi_{dia} = \psi_{\rm adv}(y_1, \sigma_\theta) - \psi_{\rm adv}(y_0, \sigma_\theta) - T_{hv} \,. \quad (6b)$$

Equation (6) is equivalent to Eq. (5) if the sloshing streamfunction is obtained by integrating southward from $\psi_{\rm adv}(y_0, \sigma_\theta)$, or equivalently northward from $\psi_{\rm adv}(y_1, \sigma_\theta)$ (i.e., Eq. (3)). Equation (6) thus clarifies why the boundary conditions for the sloshing streamfunction in Eq. (3) must be specified in terms of the advective streamfunction. The signs of $\psi_{dia}$ and $\psi_{\rm adv}$ diagnose Lagrangian and Eulerian tendencies, respectively, as summarized in Figure 1.

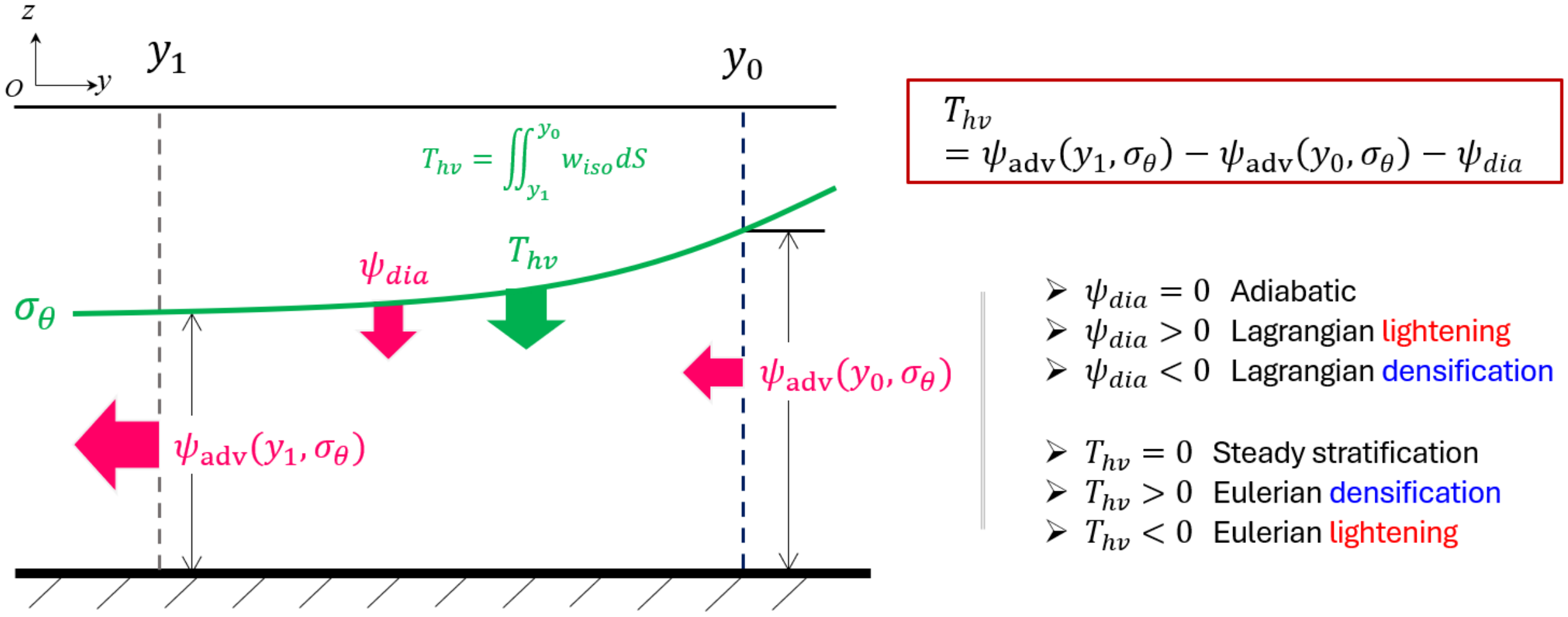

**Figure 1**. Schematic diagram of a volume budget framework for a control volume bounded by a specified isopycnal surface ($\sigma_\theta$), solid boundaries (the seafloor and eastern and western coasts), and two meridional open boundaries ($y_0$ and $y_1$). $\psi_{\rm adv}$ and $\psi_{dia}$ denote the advective and diapycnal streamfunctions in density coordinates, respectively. $T_{hv}$ represents the isopycnal volume tendency, obtained by integrating the isopycnal displacement rate ($w_{iso}$) over the horizontal area. The volume balance and the diagnostic criteria for Eulerian and Lagrangian tendencies, based on $\psi_{dia}$ and $T_{hv}$, are summarized in the right-hand column.

The three MOC streamfunctions defined above—advective, sloshing, and diapycnal—computed from monthly reanalysis fields and analyzed across three distinct regimes in the following section.

## 3. Results

### 3.1. Abyssal Indo-Pacific

#### 3.1.1. Triple-cell Lagrangian trends under Eulerian densification

Applying the decomposition framework (Eqs. 2, 3b, and 5) to the ECCO fields in the Indo-Pacific yields the time-mean advective (residual), sloshing, and diapycnal MOC streamfunctions.

A basin-wide advective upwelling cell is found in the densest part of the basin (Figure 2a) , consistent with previous studies based on ECCO state estimates (e.g., Cessi 2019; Rousselet et al. 2021; Rogers et al. 2023; Monkman and Jansen 2024).

The sloshing MOC component is also characterized by widespread upwelling, corresponding to a basin-wide upward displacement of isopycnals over the ECCO period (Figure 2b). This upward shift implies Eulerian densification (or cooling, assuming density variations are primarily temperature-controlled). In contrast, the diapycnal overturning transport ( $\psi_{dia}$ ) displays a pronounced triple-cell structure in the southern part of the basin, while remaining comparatively weak in the northern basin (Figure 2c). The weak diapycnal transport in the northern basin suggests the largely reversible nature of the upwelling, approaching the adiabatic limit ( $\psi_{dia} = 0$ ) illustrated in Figure 1.

The triple-cell diapycnal feature consists of two upwelling cells separated by an intermediate downwelling cell. The upper and lower upwelling cells are consistent with the classical view that deep-ocean waters gain buoyancy through mixing and cross isopycnals upwards. The lower cell likely reflects the transformation of Antarctic Bottom Water (AABW) driven by downward heat fluxes and geothermal heating, i.e., its consumption (de Lavergne et al 2016). The upper upwelling cell occupies a density range of $45.00$–$45.88\sigma_4$, corresponding to depths of approximately 1,000–4,000 m in ECCO, and coincides with the “middepth” Indo-Pacific upwelling (1,000–3,000 m) investigated by Rogers et al. (2023).

Together, these two diapycnal upwelling cells define a regime of Lagrangian lightening: despite the upward shift of isopycnals (Figure 2b), water parcels ascend faster than the density surfaces and are therefore transformed into lighter density classes.

However, the intermediate diapycnal downwelling is of greater significance. Although many previous studies have inferred diapycnal downwelling in regions of bottom-intensified turbulent dissipation (e.g., de Lavergne et al. 2016; Ferrari et al. 2016; McDougall and Ferrari 2017), it has not yet been explicitly diagnosed from observationally constrained datasets. Here, we present the first such diagnosis, underscoring the utility of the proposed framework.

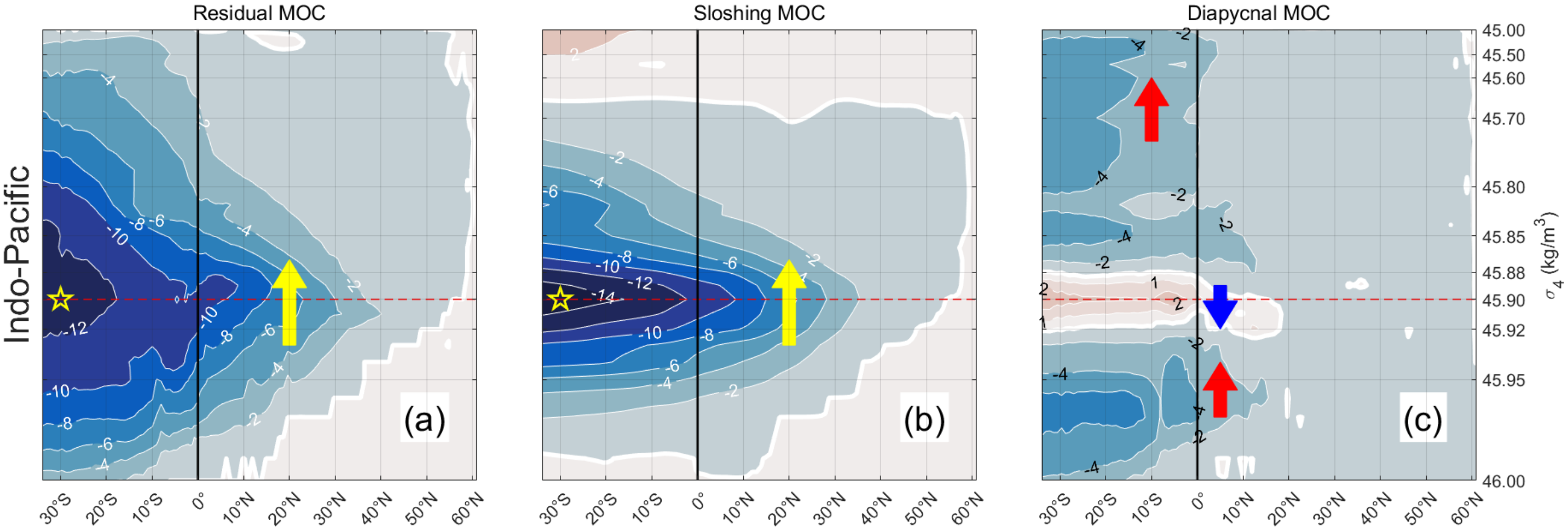

**Figure 2**. Time-mean MOC streamfunction components of the Indo-Pacific Ocean below approximately 1,000 m in density ($\sigma_4$) coordinates. (a) Residual MOC streamfunction ($\psi_{res}$); (b) Sloshing MOC streamfunction ($\psi_{slo}$); (c) Diapycnal or diffusive MOC streamfunction ($\psi_{dia}$). Positive values (warm colors) denote a clockwise overturning cell (downwelling to the north of the maximum), while negative values (cool colors) denote anticlockwise overturning cell (upwelling to the north). Vertical arrows indicate the sense of overturning in each panel. Yellow stars mark the locations at which the MOC indices are defined. Data: ECCO v4r3. Units: Sv.

Figure 2 also shows that the vertical volume transport near the residual overturning maximum is dominated by the sloshing mode. In other words, water parcels in this region predominantly upwell in an adiabatic manner, with material surfaces rising largely in phase with density surfaces. It highlights the reversibility of this Eulerian densification.

This physical linkage is further supported by their temporal covariation. An overturning index is defined for both streamfunctions as the value near the maximum (30°S, 45.90 $\sigma_4$). The resulting time series, smoothed using 1-year and 5-year windows, respectively, are shown in Figure 3. Strong correlations of 0.92 and 0.88 are obtained at the respective timescales, with the smoothed time series detrended prior to the correlation analysis.

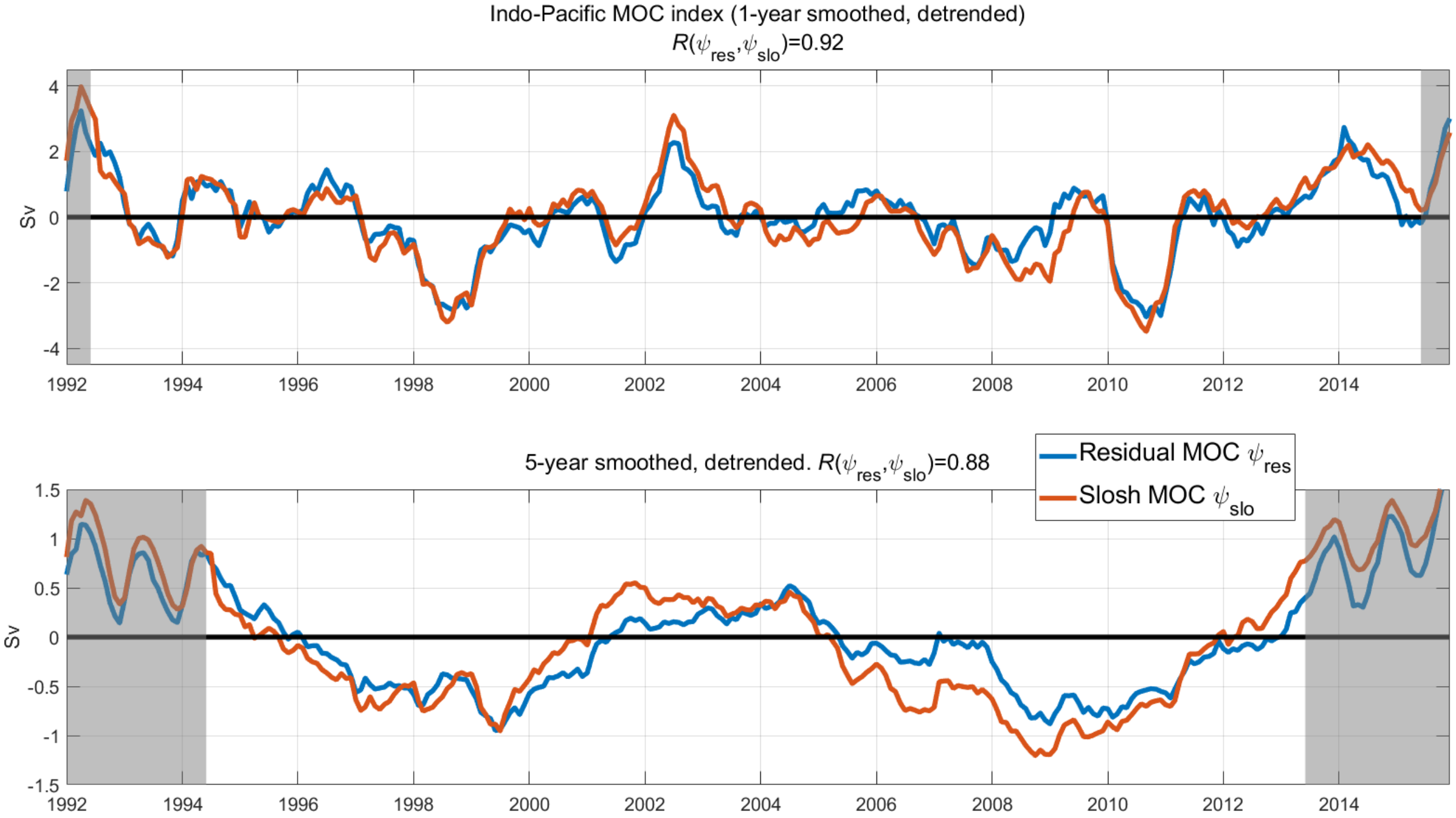

**Figure 3**. Interannual time series of the MOC indices at 30°S and 45.90$\sigma_4$ (yellow star in Figure 2), showing the residual (blue) and sloshing (red) components. To minimize edge effects associated with incomplete averaging windows, the first and last $N/2$ data points (grey shaded; $N$ = 12 and 60 for the 1-year and 5-year filters, respectively) were excluded from the correlation analysis.

### 3.1.2. Co-existing diapycnal downwelling and Eulerian upwelling

In contrast, the intermediate diapycnal downwelling—i.e., Lagrangian densification—spanning a density range of approximately 45.88–45.92 $\sigma_4$ (depths of roughly 3,500–4,500 m), emerges as a striking and previously underrecognized feature. It points to an unexpected regime in which Eulerian upwelling coexists with diapycnal downwelling; that is, water parcels ascend in physical space while simultaneously crossing isopycnals toward denser classes. This apparently paradoxical behavior is of particular importance, as it provides new insights into the long-standing closure problem of the abyssal overturning circulation.

Munk's abyssal upwelling theory, well known as the "abyssal recipes" (Munk 1966), predicts a net diapycnal downwelling rather than upwelling in the ocean interior when subject to observed bottom intensified turbulent dissipation (e.g., St. Laurent et al. 2001; Ferrari et al. 2016; Callies and Ferrari 2018; Drake et al. 2020). This inferred downwelling is in apparent contradiction with the requirement of mass balance in the deep ocean. To reconcile this inconsistency, previous resolutions have invoked upslope transport within the bottom boundary layer along the seafloor (e.g., McDougall and Ferrari, 2017).

Here, however, our analysis reveals an alternative dynamical regime in which diapycnal downwelling does not preclude Eulerian upwelling. Instead, water parcels in the ocean interior can continue to upwell while undergoing diapycnal transformation toward denser classes. This process of Lagrangian densification has been conceptually illustrated in Fig. 4b of Han (2021), and is here diagnosed explicitly as a dynamically significant component of the overturning circulation. This decoupling between Eulerian and Lagrangian perspectives offers a new pathway for reconciling bottom-intensified mixing with the large-scale overturning mass balance.

This diapycnal upwelling cell agrees with the volume-budget analysis presented by Monkman and Jansen (2024). Their results show a meridional transport convergence of 14.5 Sv and a volume tendency of 17.0 Sv in the Indo-Pacific basin, both peaking at 37.053 $\sigma_2$, i.e., near the maximum of the advective overturning, as also shown in Figure 2. According to the framework illustrated in Figure 1, these values imply a diapycnal downwelling of 2.5 Sv. Nevertheless, Monkman and Jansen (2024) interpret the 14.5 Sv transport convergence as "diapycnal upwelling", reflecting a difference in diagnostic interpretation, rather than a fundamental discrepancy in the underlying circulation. This distinction is discussed further in Section 3.2.2.

### 3.1.3. Multiple lines of evidence for Eulerian and Lagrangian densification signals

Based on the proposed decomposition, two prominent trends emerge in the abyssal Indo-Pacific: a basin-wide Eulerian densification and a layer of Lagrangian densification. To support these findings, we draw on multiple lines of evidence, including observational constraints and independent reanalysis-based diagnostics.

Beyond the cooling trend identified in the same dataset by earlier studies (Wunsch and Heimbach 2014; Liang et al. 2015; Monkman and Jansen 2024), the widespread Eulerian densification signal in the Indo-Pacific basin is also supported by independent observations. A comparison between HMS *Challenger* observations from the 1870s and modern hydrography reveals a centennial cooling of approximately 0.02°C in the deep Pacific, a trend that intensifies with depth (Gebbie and Huybers 2019). Furthermore, microstructure measurements consistently indicate bottom-intensified vertical diffusive buoyancy fluxes (e.g., Polzin et al. 1996; Morris et al. 2001; St. Laurent et al. 2001; Waterman et al. 2013; Mashayek et al. 2017), a feature dynamically consistent with the inferred Eulerian densification.

At the same time, modern observational constraints on abyssal trends remain limited and sometimes contradictory. Deep Argo and historical ship-based observations from the mid-1980s to the mid-2010s reveal spatially heterogeneous warming and cooling signals in the deep Indo-Pacific (2,000–4,000 dbar), alongside a more consistent warming trend in the abyssal layer (4,000–6,000 dbar). These discrepancies underscore the ongoing challenge of robustly constraining long-term abyssal changes from direct observations alone.

In comparison, the Lagrangian densification trend identified near 45.90 $\sigma_4$ is inherently difficult to confirm directly from observation. Nevertheless, it is consistent with the results of Monkman and Jansen (2024), who identified regimes where the isopycnal volume tendency exceeds the Eulerian volume transport by 2.5 Sv near the maximum of the abyssal overturning cell in the Indo-Pacific at 30°S, as shown in Figure 2c.

Finally, it is important to note that ECCO likely underestimates diapycnal mixing in the deep ocean compared to independent estimates based on microstructure observations and internal wave–driven mixing (Trossman et al. 2022). This bias may distort the balance between bottom-water intrusion and mixing-induced transformation, thereby affecting the transient state of the abyssal ocean and biasing the diagnosed diapycnal overturning. In particular, the relatively low diffusivity in ECCO, corresponding to weaker mixing, may lead to an overestimation of diapycnal downwelling in the abyssal Indo-Pacific.

Despite these uncertainties, the sloshing solution (Figure 2b) provides a dynamically consistent constraint on Eulerian tendencies, as it directly reflects isopycnal displacement and is not contingent on the specific limitations of the underlying dataset.

### 3.2. Atlantic: similar Lagrangian trends despite contrasting Eulerian lightening

Unlike the Indo-Pacific, time-mean residual overturning in the Atlantic exhibits a well-defined clockwise overturning cell (Figure S1 in the Supplementary Material), consistent with previous studies (e.g., Johnson et al. 2019; Rousselet and Cessi 2022; Monkman and Jansen 2024). The AMOC streamfunction is computed up to 60°N, beyond which it becomes ill-defined due to open zonal boundaries (Monkman and Jansen 2024).

However, the strong mean state of the AMOC can obscure its meridional evolution. To better isolate the latitudinal variation of the overturning, we define a streamfunction anomaly relative to the northern boundary by subtracting the profile at 60°N from all latitudes. This anomaly representation effectively removes the dominant mean signal while highlighting the latitudinal structure of the MOC. The diapycnal MOC remains unaffected by this operation, as the sloshing and residual streamfunctions share the same boundary condition (see discussion of Eq. 6). The resulting fields are shown in Figure 4.

As revealed by the sloshing MOC anomaly (Figure 4b), the deep and abyssal Atlantic exhibits a basin-wide Eulerian lightening tendency, in contrast to the Indo-Pacific. Despite this contrasting Eulerian signal, the Lagrangian tendency inferred from the diapycnal overturning exhibits a remarkably similar triple-cell structure to that in the Indo-Pacific, with the primary difference being a weaker intermediate diapycnal downwelling cell (Figure 4c).

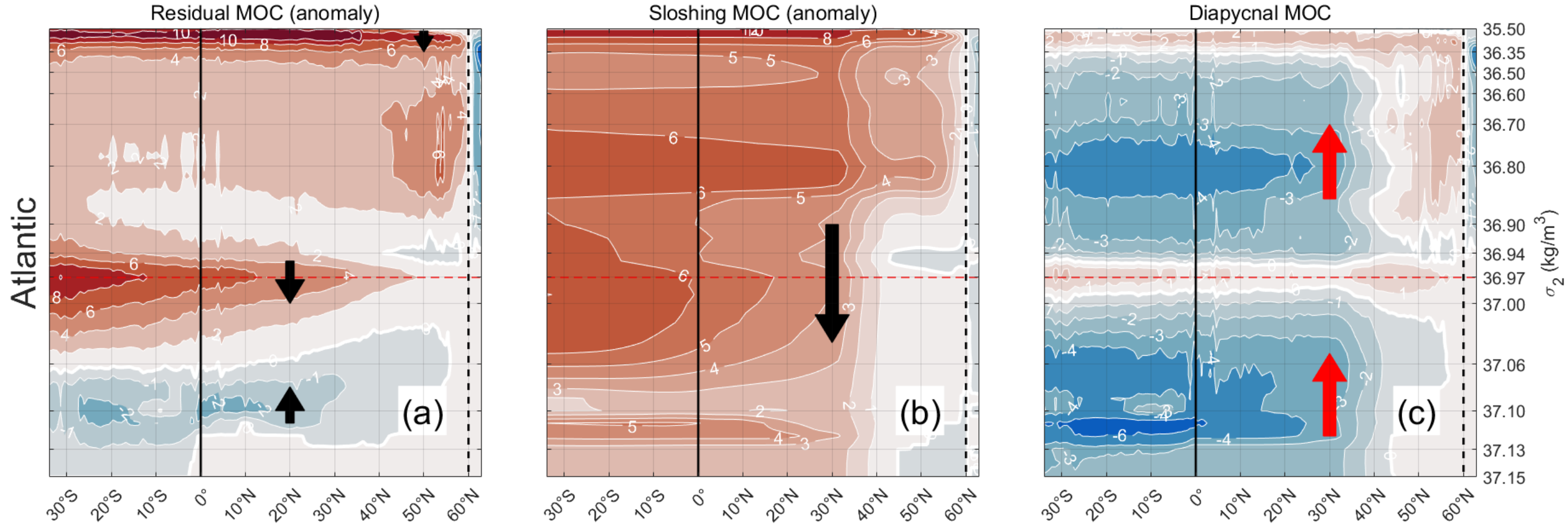

**Figure 4**. Dynamical overturning decomposition for the Atlantic. The panel layout follows Figure 2, except that $\sigma_2$ coordinates are used. For the residual and sloshing MOC, streamfunction anomalies relative to their profiles at 60°N (black dashed line) are shown.

### 3.2.1. The diapycnal downwelling cell under Eulerian lightening

Although the triple-cell diapycnal overturning structure in the Atlantic closely resembles that in the Indo-Pacific, a key dynamical difference emerges. In the Atlantic, the Lagrangian densification cell, centered around 36.97 $\sigma_2$, coincides with Eulerian downwelling, rather than Eulerian upwelling (Figure 4).

This regime has not been included in the conceptual diagram in Fig. 4 of Han (2021), and is therefore schematically represented here in Figure S2. Physically, this behavior arises when Eulerian downwelling is accompanied by a densification process (e.g., cooling associated with bottom-intensified turbulent heat flux), such that water parcels are transformed into denser classes while descending.

As in the Indo-Pacific, this diapycnal downwelling transport is consistent with the volume budget analysis of the Atlantic by Monkman and Jansen (2024). Their results show an isopycnal volume tendency of approximately −6 Sv (their *dV/dt*) and a meridional advective transport convergence (their $\Delta\Psi$) at 32°S of about −9 Sv, peaking at around 35.96 $\sigma_2$ (their Figure 4a), in good agreement with the structure shown in Figure 4 or Figure S1.

### 3.2.2. Widespread Eulerian and Lagrangian lightening in the deep Atlantic

The widespread Eulerian lightening is consistent with the significant warming trend between 500 m and 3,000 m of the Atlantic reported by Gebbie and Huybers (2019), but is less consistent with the findings of Johnson and Purkey (2024) based on modern observations.

The anomalous residual streamfunction exhibits a broad domain of weak meridional variation within the density range 35.50−36.90 $\sigma_2$ up to 40°N, implying minimal convergence of meridional transport within this layer. This feature coincides with the middepth diapycnal upwelling cell centered near 36.80 $\sigma_2$. Within this regime, the overturning can be interpreted using the framework illustrated in Figure 1: when $\psi_{\mathrm{adv}}(y_1, \sigma_\theta) \approx \psi_{\mathrm{adv}}(y_0, \sigma_\theta)$, Eq. (6) reduces to

$$\psi_{dia} \approx -T_{hv} \,. \tag{7}$$

This relationship indicates that the diapycnal upwelling in this regime is primarily associated with the downward displacement of isopycnals, which is concentrated north of 30°N. It further highlights that the advective (Eulerian or residual) overturning streamfunction alone does not adequately reflect diapycnal transport.

This interpretation differs slightly from that of Monkman and Jansen (2024), who interpret the meridional transport convergence in density space (their $\Delta\Psi$) as the "net diapycnal transport at each density surface". In our framework (Figure 1), the diapycnal transport is instead expressed as $\Delta\Psi - T_{hv}$, rather than $\Delta\Psi$ alone. Accordingly, we interpret their reported close correspondence between $\Delta\Psi$ and $T_{hv}$ as indicating a leading role of the isopycnal volume tendency ($T_{hv}$) in shaping advective transport divergence, rather than in diabatic water-mass transformations.

Beyond this middepth regime, the abyssal diapycnal upwelling cell ($\sigma_2$>36.97) reflects a combined influence of isopycnal deepening and Eulerian upwelling, as evidenced by the bottom upwelling cell in Figure 4a.

### 3.2.3. A sub-overturning cell in the upper limb of the AMOC

An upper-ocean overturning anomaly cell is identified, although it is partially obscured in Figure 4 due to the use of a bottom-stretched density coordinate. When viewed in a non-stretched density coordinate (Figure 5), this feature emerges clearly as a pronounced clockwise residual overturning cell. For reference in physical space, the zonally averaged, time-mean Atlantic density structure corresponding to the density range of the sub-cell is shown in Figure S3.

This clockwise residual overturning cell is primarily confined to depths shallower than 1,000 m, i.e., within the upper limb of the AMOC. The relatively small difference between the residual and sloshing MOC (Figure 5c) indicates that this anomaly cell is largely associated with downward displacement of isopycnals. Both sub-cells are predominantly located north of 45°N, within the downwelling region (highlighted by yellow boxes in Figure 5), where their streamlines are closely aligned (highlighted in blue). The tilt of these downwelling streamlines suggests that denser isopycnals descend more rapidly than lighter ones. This vertical divergence in isopycnal volume must be compensated by horizontal convergence, which, under potential vorticity conservation, implies a poleward transport (i.e., stretching of water columns).

Although the causality between horizontal transport convergence and vertical isopycnal divergence remains uncertain, one plausible mechanism involves the formation of intermediate waters in the subpolar North Atlantic, such as Subarctic Intermediate Water (SAIW). This water mass originates from surface waters at 50–60°N and occupies depths of approximately 100–500 m (Liu and Tanhua 2021). Its formation acts as a volumetric source within the corresponding density classes, increasing the volume of isopycnal layers. Through potential vorticity conservation, the associated column stretching would in turn drive a northward transport. In analogy to the role of North Atlantic Deep Water (NADW) formation in sustaining the AMOC, the formation of intermediate waters may likewise contribute to the development of this sub-overturning cell in the upper 1,000 m of the Atlantic. This points to a dynamically analogous, albeit weaker, overturning pathway operating in the upper limb of the AMOC.

These cells have no direct counterpart in the volume budget analysis of Monkman and Jansen (2024), as this density range was not considered in their study. Additionally, the correspondence of the two sub-cells in the yellow boxes of Figure 5 is less clear for the Eulerian MOC streamfunction. This indicates that parameterized eddies associated with the tilted isopycnals (Figure S3) play an important role in the overturning transport in this region.

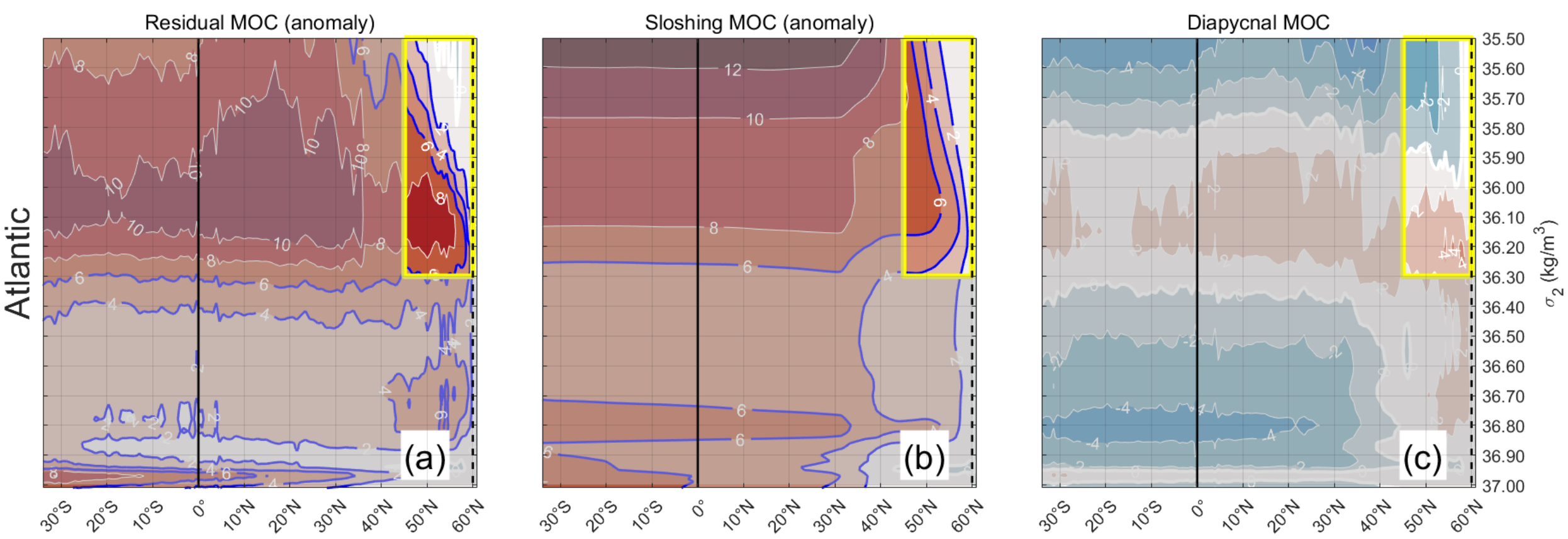

**Figure 5**. Same as Figure 4, but shown in a non-stretched density coordinate to highlight the sub-overturning cell in the upper limb of the AMOC. Yellow boxes indicate the primary downwelling region (50–60°N and 35.50−36.30 $\sigma_2$). Selected downwelling streamlines (2, 4, and 6 Sv) are highlighted by blue contours.

### 3.3. Deep South China Sea: Eulerian densification masking Lagrangian lightening

Moving beyond the major open-ocean basins, we extend our decomposition framework to the South China Sea (SCS), a semi-enclosed marginal sea that connects the Pacific and Indian Oceans. Owing to its complex topography and restricted geometry, the SCS provides an ideal natural laboratory for investigating basin exchange and interior overturning dynamics. While the central basin exceeds depths of 4,000 m, its only deep connection to the open ocean is through the Luzon Strait (Tian et al. 2006; Yang et al. 2010; Zhao et al. 2014). Consequently, the deep SCS effectively functions as a closed reservoir below the sill depth, ventilated primarily by the cascading inflow of dense Pacific water through the strait (Zhao et al. 2014; Zhou et al. 2014). This inflow, driven by the density gradient across the strait (Zhou et al. 2014), sustains a deep overturning cell that is conventionally assumed to be balanced by mixing-induced upwelling (Wang et al. 2016; Cai and Gan 2020). Evidence for this overturning has been largely inferred from geochemical tracers, particularly dissolved oxygen, which distinguish aged interior waters from the younger, oxygen-rich Pacific source (Chen et al. 2001; Li and Qu 2006; Li et al. 2017). However, due to the scarcity of direct deep observations, our understanding of these processes still relies heavily on numerical models and reanalyses (e.g., Xie et al. 2013; Shu et al. 2014; Gan et al. 2016; Zhu et al. 2016, 2019; Li et al. 2025).

Given the limited spatial extent of the SCS (approximately 12°×12°), the coarse 1° resolution of ECCO is inadequate. We therefore employ the eddy-resolving 1/12° GLORYS reanalysis (Section 2.1.2). Geometrically, the deep SCS basin is bounded to the south by a closed wall boundary and opens northward to the Pacific via the Luzon Strait; accordingly, Eq. (3a) is adopted to construct the sloshing streamfunction. For the advective overturning circulation, the Eulerian streamfunction computed using Eq. (1) is adopted, as mesoscale eddies are largely resolved at this resolution.

Figure 6 shows the time-mean Eulerian, sloshing, and diapycnal MOC streamfunctions averaged over the GLORYS period (1993–2024) in $\sigma_2$ coordinate. Solutions within the Luzon Strait are masked due to the presence of an open zonal boundary. The Eulerian overturning exhibits a canonical structure, with upwelling in the southern basin and downwelling in the north (Figure 6a), consistent with previous studies (e.g., Wang et al. 2016; Zhu et al. 2016; Wang et al. 2017) and reflecting the intrusion of denser Pacific water followed by interior upwelling.

The clockwise sloshing MOC cell reveals a basin-wide upward displacement of deep isopycnals (Figure 6b), corresponding to a pervasive Eulerian densification signal. Notably, this finding is consistent with the observational estimates indicating a cooling trend in the 2,000–4,000 m layer of the SCS (Johnson and Purkey 2024), and is dynamically consistent with cooling in the deep Pacific (Gebbie and Huybers 2019). As the deep SCS is directly ventilated by Pacific inflow, this Eulerian densification can be interpreted as a regional imprint of broader open-ocean tendency.

In contrast, the diapycnal overturning cell is substantially weaker than the advective and sloshing components, particularly in the upwelling regime (Figure 6c), indicating strong cancellation between these cells. This suggests that bottom-water upwelling in the SCS is predominantly adiabatic: water parcels are mechanically lifted by the continuous intrusion of dense Pacific inflow, while diabatic mixing is insufficient to offset this supply in time and sustain a steady stratification.

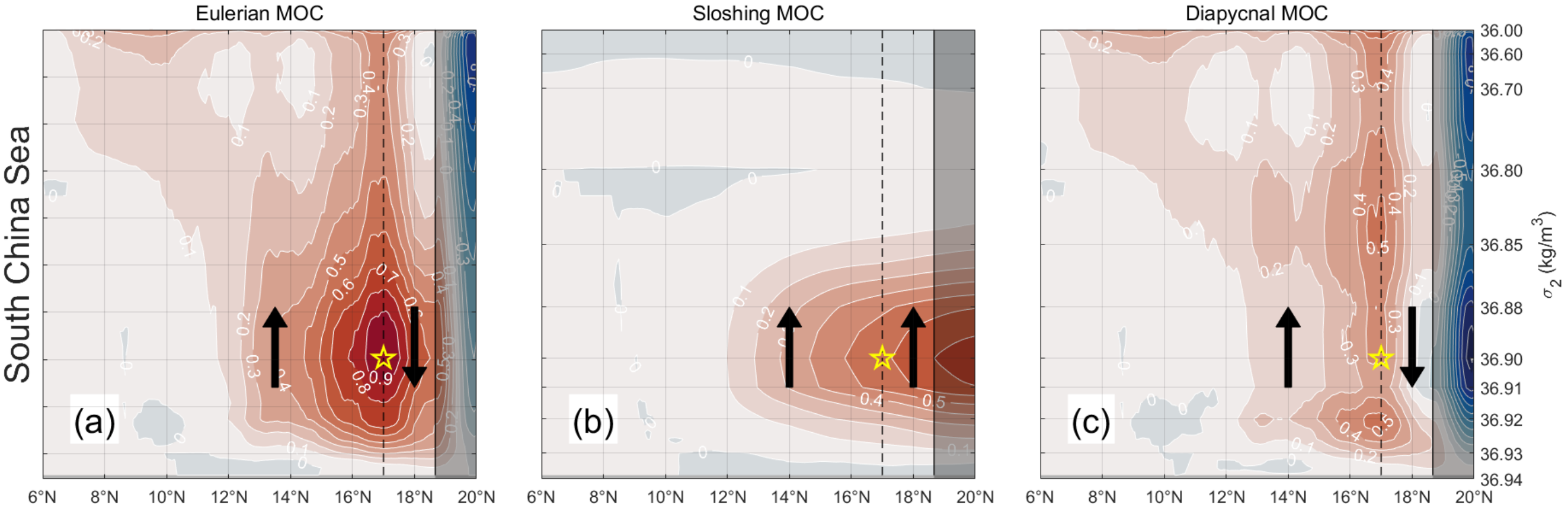

**Figure 6**. Dynamical overturning decomposition for the deep South China Sea (below approximately 700 m) based on GLORYS. Panels follow the layout of Figure 2. Solutions within the Luzon Strait are masked due to the presence of an open zonal boundary. The Eulerian MOC streamfunction is meridionally smoothed using a 1° window to suppress high-frequency noise, whereas no smoothing is applied to the sloshing streamfunction. The yellow star marks the location of the MOC index.

As for the Indo-Pacific, the linkage between the Eulerian upwelling and isopycnal heaving processes can be further assessed by their temporal covariation. A deep MOC index is defined near the Eulerian maximum (17°N, 36.90$\sigma_2$; yellow star in Figure 6). In the sloshing time series, sporadic abrupt changes in isopycnal depth are detected (see an example in Figure S4), which can contaminate the index. These outliers (~2.6% of the data) are identified using a median absolute deviation (MAD) criterion, with a threshold of 3×MAD relative to a 12-month moving median, and are subsequently removed.

The cleaned and detrended time series (Figure 7) exhibit strong correlations (R = 0.81 and 0.88 at one-year and five-year timescales, respectively), indicating a tight coupling between advective and sloshing transports in the deep SCS. This behavior is consistent with findings in the Atlantic, where sloshing variability dominates the overturning signal.

In summary, the GLORYS reanalysis indicates that adiabatic isopycnal heaving, rather than diabatic mixing, primarily governs the Eulerian tendency in the deep SCS: the rate of external replenishment via Pacific inflow exceeds the rate of internal diabatic consumption. In sharp contrast, the Lagrangian perspective reveals a net lightening tendency south of 17°N, indicating that bottom waters are progressively transformed into lighter density classes during their ascent.

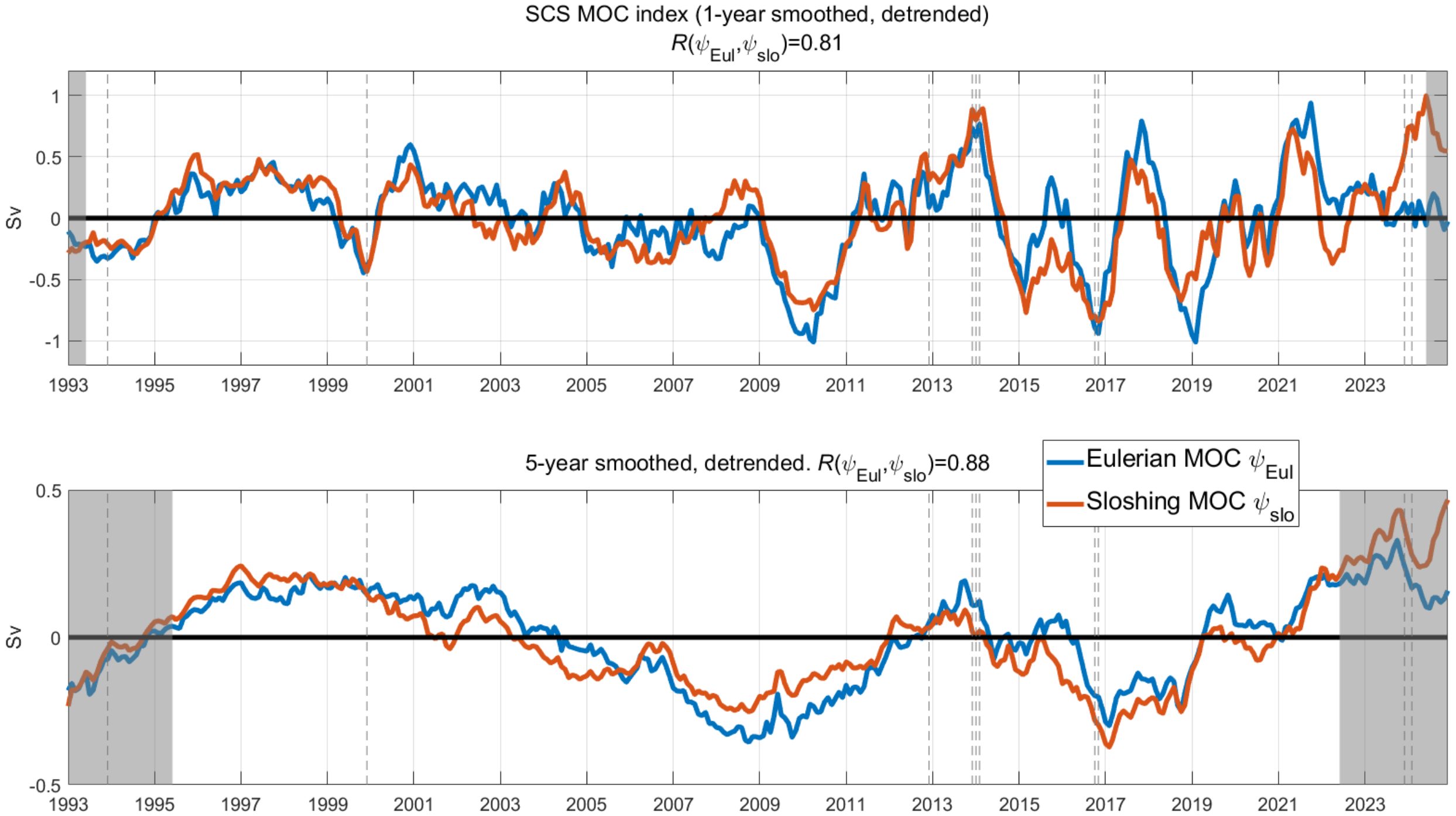

**Figure 7**. Same as Figure 3, but for the deep SCS MOC indices at 17°N and 36.90$\sigma_2$ (yellow star in Figure 6) Months identified as outliers in the sloshing index are indicated by vertical dashed lines.

## 4. Discussion and summary

Distinguishing between Eulerian and Lagrangian tendencies is fundamental to understanding ocean processes and their changes, such as heat storage, advective transport, and water-mass transformations; yet, these two perspectives are rarely separated in standard diagnostics. By applying a dynamical overturning decomposition framework to two reanalysis products, we show that Eulerian tendencies and Lagrangian transformations in the ocean interior can be effectively disentangled. This separation provides a dynamical basis for distinguishing between reversible and irreversible changes in the ocean interior, with direct implications for interpreting long-term climate signals.

The utility of this framework is demonstrated across three contrasting dynamical regimes: the Indo-Pacific, the Atlantic, and the South China Sea (SCS). The corresponding Eulerian and Lagrangian tendencies are summarized in Table 1.

A central insight emerging from this analysis is that Eulerian tendencies can mask the underlying Lagrangian evolution, as the two are governed by distinct physical processes. The primary conceptual advance of this study is the explicit quantification of Lagrangian tendencies, whose magnitude relative to advective transport provides a direct measure of the reversibility of Eulerian tendencies.

**Table 1**. Eulerian and Lagrangian tendencies in terms of potential density for the three examined basins.

| Basin | Vertical volume transport | Eulerian tendency | Lagrangian tendency | Dataset |
|---|---|---|---|---|
| **Indo-Pacific** (below ~1,000 m) | Upwelling | Densification (isopycnal upwelling) | Lightening and densification (diapycnal upwelling and downwelling) | ECCO (1°) |
| **Atlantic** | Deep: downwelling; Abyss: upwelling | Lightening (isopycnal downwelling) | Lightening (diapycnal upwelling) | ECCO (1°) |
| **South China Sea** (below ~700 m) | Upwelling | Densification (isopycnal upwelling) | Lightening (diapycnal upwelling) | GLORYS (1/12°) |

Beyond this general key message, several specific findings from the three basins examined emerge from the decomposition analysis.

- *Case 1 (Indo-Pacific)*: The basin-wide diapycnal upwelling is interrupted by an intermediate diapycnal downwelling cell. The identification of this diapycnal downwelling cell is of particular importance for the theoretical study of abyssal circulation for two reasons. First, although previous studies have inferred diapycnal downwelling from bottom-intensified mixing, it has not yet been explicitly diagnosed from observationally constrained datasets, representing the first direct diagnosis of such a process. Second, it may help reconcile the apparent closure paradox between Munk's abyssal upwelling and bottom-intensified turbulent buoyancy fluxes, which favor diapycnal downwelling (e.g., Ferrari et al. 2016). Specifically, it demonstrates that diapycnal downwelling and Eulerian upwelling are not mutually exclusive: water parcels may undergo Eulerian ascent while simultaneously experiencing downward diapycnal transport.

With regard to the diapycnal upwelling cells, it is important to note that they are not associated with the bottom boundary layer (BBL) upwelling proposed in previous studies (e.g., Ferrari et al., 2016; McDougall and Ferrari, 2017). According to the theory, such upwelling is typically confined to a thin and spatially localized BBL, with a horizontal scale of order ~0.2° and a thickness of ~50 m (McDougall and Ferrari, 2017), which cannot be resolved at the coarse spatial resolution of ECCO.

Additionally, north of 20°N, the much weaker diapycnal transport suggests that Eulerian upwelling occurs in a more nearly adiabatic manner than the southern part of the basin (30°S–20°N).

- *Case 2 (Atlantic)*: The diapycnal overturning in the Atlantic exhibits a triple-cell structure broadly similar to that in the Indo-Pacific, despite a contrasting Eulerian background of widespread lightening. In addition, two clockwise overturning sub-cells are identified within the AMOC. The lower sub-cell, located within the lower limb of the AMOC, coincides with the diapycnal downwelling cell. The upper sub-cell extends from the depth of ~1,000 m to its surface-outcropping region in the subpolar North Atlantic, i.e., within the upper limb of the AMOC. This upper sub-cell is plausibly linked to the formation of intermediate water masses, such as SAIW, whereas the mechanism of the lower sub-cell remains unclear. The presence of these sub-cells may highlight a previously overlooked structural feature of the AMOC.
- *Case 3 (South China Sea)*: The SCS features a regime in which Eulerian densification coexists with Lagrangian lightening. Our decomposition reveals that the observed deep densification in the SCS is primarily driven by adiabatic heaving forced by Pacific inflow through the Luzon Strait, thereby masking the comparatively weaker background of diabatic mixing. This result highlights that similar or even identical Eulerian tendencies can correspond to fundamentally different Lagrangian processes: Lagrangian densification in the abyssal Indo-Pacific contrasts with Lagrangian lightening in the deep SCS.

None of the three study cases clearly represents a state of steady stratification, as identified as one possible regime in Figure 1. This is unsurprising, because reanalysis products spanning a limited time period inevitably reflect transient states, characterized by either densification or lightening. Under such circumstances, numerical modeling becomes essential for exploring the equilibrium status of the overturning circulation. In an idealized model configuration integrated over 4,000 years, we obtain a quasi-equilibrium solution (manuscript in preparation), in which the sloshing component is negligible almost everywhere, such that the Eulerian upwelling largely accounts for the diapycnal transport, corresponding to the limit of $T_{hv} = 0$ in Figure 1.

Further support for the framework comes from the strong correlations between the advective and sloshing overturning indices in the Indo-Pacific and SCS, as well as the Atlantic (Han et al., 2023a). This provides a useful internal consistency check because the correlated indices are generated from two physically distinct fields: meridional velocity and density. Their close agreement suggests that the framework identifies coherent and dynamically consistent processes.

This study differs from previous work using related diagnostics in two important aspects. First, although it adopts the same overturning decomposition approach as in previous studies by the author (e.g., Han 2021, 2023a), the present work focuses on mean states and long-term tendencies, rather than short-term variability, with a particular emphasis on diagnosing *diapycnal* transport, which was not a primary focus of our earlier work. Second, the "isopycnal overturning" of Monkman and Jansen (2024) is conceptually equivalent to the "sloshing overturning" introduced in Han (2021). While their study likewise examined long-term trends in the abyssal Indo-Pacific and emphasized the dominant role of isopycnal volume tendency in abyssal upwelling, our interpretation differs slightly in the diagnosis of diapycnal transport, as discussed in Section 3.2.2. This distinction is physically important, because identification of the diapycnal downwelling cell in the Indo-Pacific points to a potentially significant addition to abyssal circulation theory and offers fresh insight into the long-standing Munk interior-downwelling conundrum.

We acknowledge that the results presented here depend on the fidelity of the reanalysis products employed. Despite the scarcity of direct observations, several key features identified here are supported by independent observation-based constraints. The coherent signals across multiple basins therefore suggest that the decoupling between Eulerian and Lagrangian tendencies reflects a robust and physically meaningful aspect of ocean dynamics. While quantitative details may remain uncertain, the qualitative structures revealed here point to previously underappreciated dynamical regimes in ocean overturning.

As global observing systems expand—particularly with the sustained development of Deep Argo—and reanalysis products become increasingly reliable in the historically data-sparse regions, this framework should enable progressively more robust diagnoses of ocean interior tendencies. This capability is especially timely given the broader climate context. For instance, while the IPCC AR6 projects that the AMOC "will very likely decline over the 21st century" according to climate modeling (Fox-Kemper et al. 2021), a key unresolved question is whether such large-scale trends reflect reversible adiabatic adjustment or irreversible diabatic transformation. By offering a dynamical framework for separating these components, our approach not only refines the interpretation of observed and modeled climate signals, but also advances a more mechanistic understanding of deep-ocean overturning in a changing climate.

**Acknowledgments**

This work was supported by NSFC grant 42476006 and Xiamen University Malaysia Research Fund grant XMUMRF/2024-C14/ICAM/0017.

**Data Availability Statement**

The ECCO reanalysis dataset (Version 4, Release 3) is openly available at https://ecco.jpl.nasa.gov.The GLORYS output is available at https://doi. org/10.48670/moi-00021. The Gibbs-Seawater Oceanographic Toolbox (McDougall and Barker 2011) software was also used and is available at https://www.teos-10.org/software.htm.

# Differentiating Eulerian and Lagrangian Tendencies in the Ocean Interior via a Dynamical Overturning Decomposition

Lei Han [a, b]

[a] *China-ASEAN College of Marine Sciences, Xiamen University Malaysia, Sepang, Malaysia*

[b] *College of Ocean and Earth Sciences, Xiamen University, Xiamen, China*

*Corresponding author*: Lei Han, lei.han@xmu.edu.my

## Supplemental materials

This document includes

- Four supplemental figures, from Figure S1 to Figure S4;

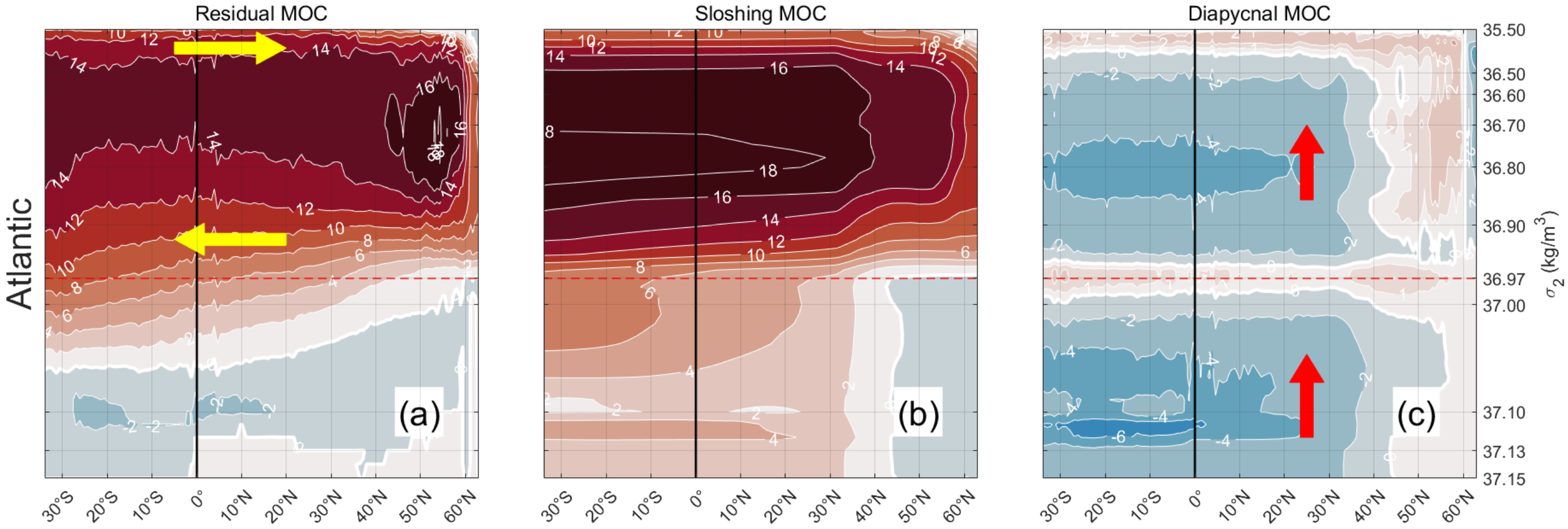

Figure S1. Time-mean residual (a), sloshing (b), and diapycnal (c) MOC streamfunctions for the Atlantic basin, analogous to the Indo-Pacific solutions shown in Figure 1. Data: ECCO v4r3.

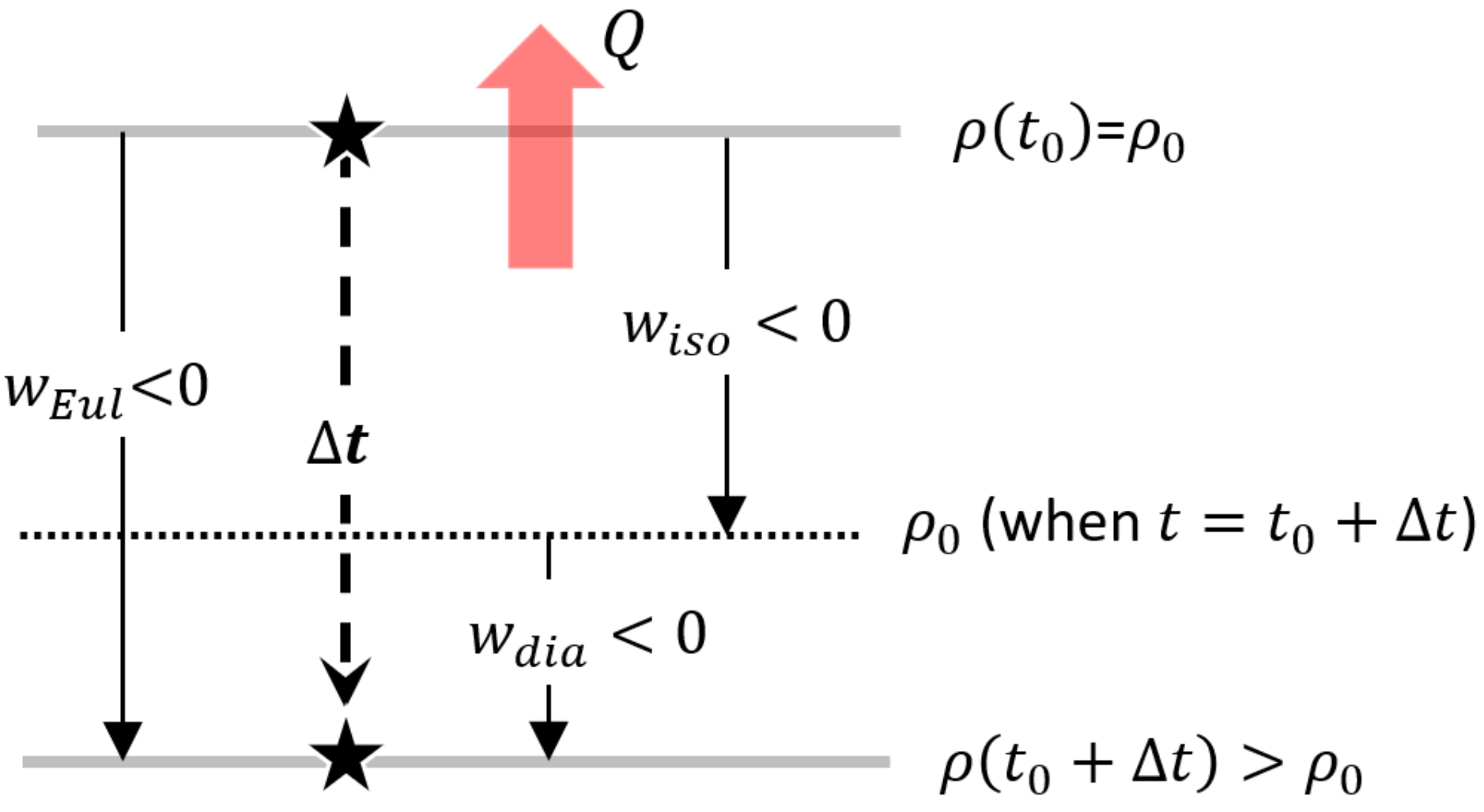

Figure S2. Schematic diagram illustrating the relationship among vertical velocities in the Eulerian framework, following Han (2021), but adapted for the case of Eulerian *downwelling* under Lagrangian cooling. The vertical velocities include the Eulerian vertical velocity ($w_{Eul}$), the isopycnal displacement rate ($w_{iso}$), and their residual, the diapycnal vertical velocity ($w_{dia}$). All notations follow Fig. 4 of Han (2021): gray solid lines denote isopycnals, the star marks a water parcel, and the thick dashed line indicates the updated position of isopycnal $\rho_0$ as time evolves from $t_0$ to $t_0 + \Delta t$, under the combined effects of Eulerian movement and heat flux.

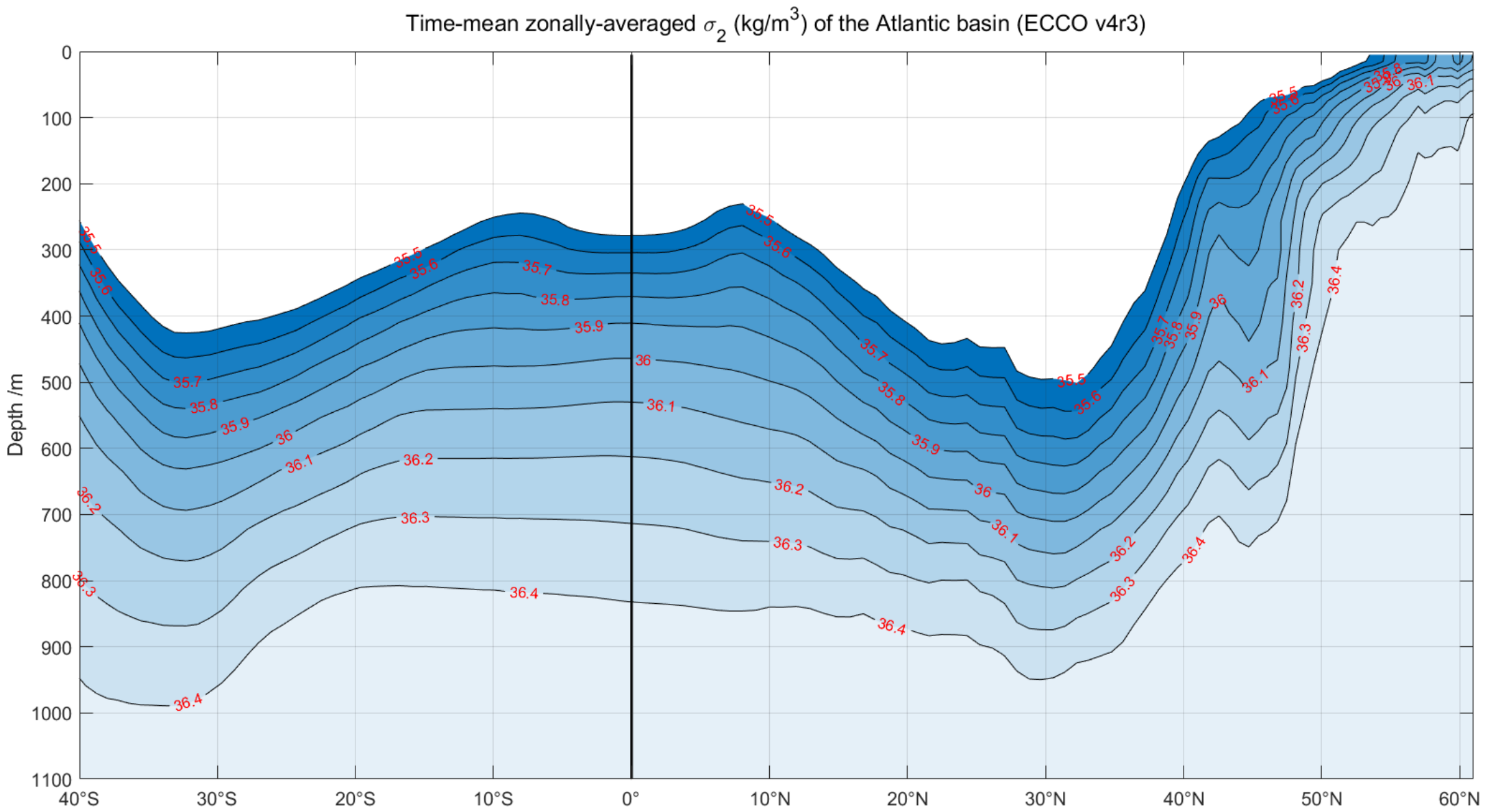

Figure S3. Time-mean, zonally averaged potential density ($\sigma_2$, referenced to 2,000 dbar) in the upper Atlantic. Only isopycnals in the range $\sigma_2$ = 35.50–36.40 are shown.

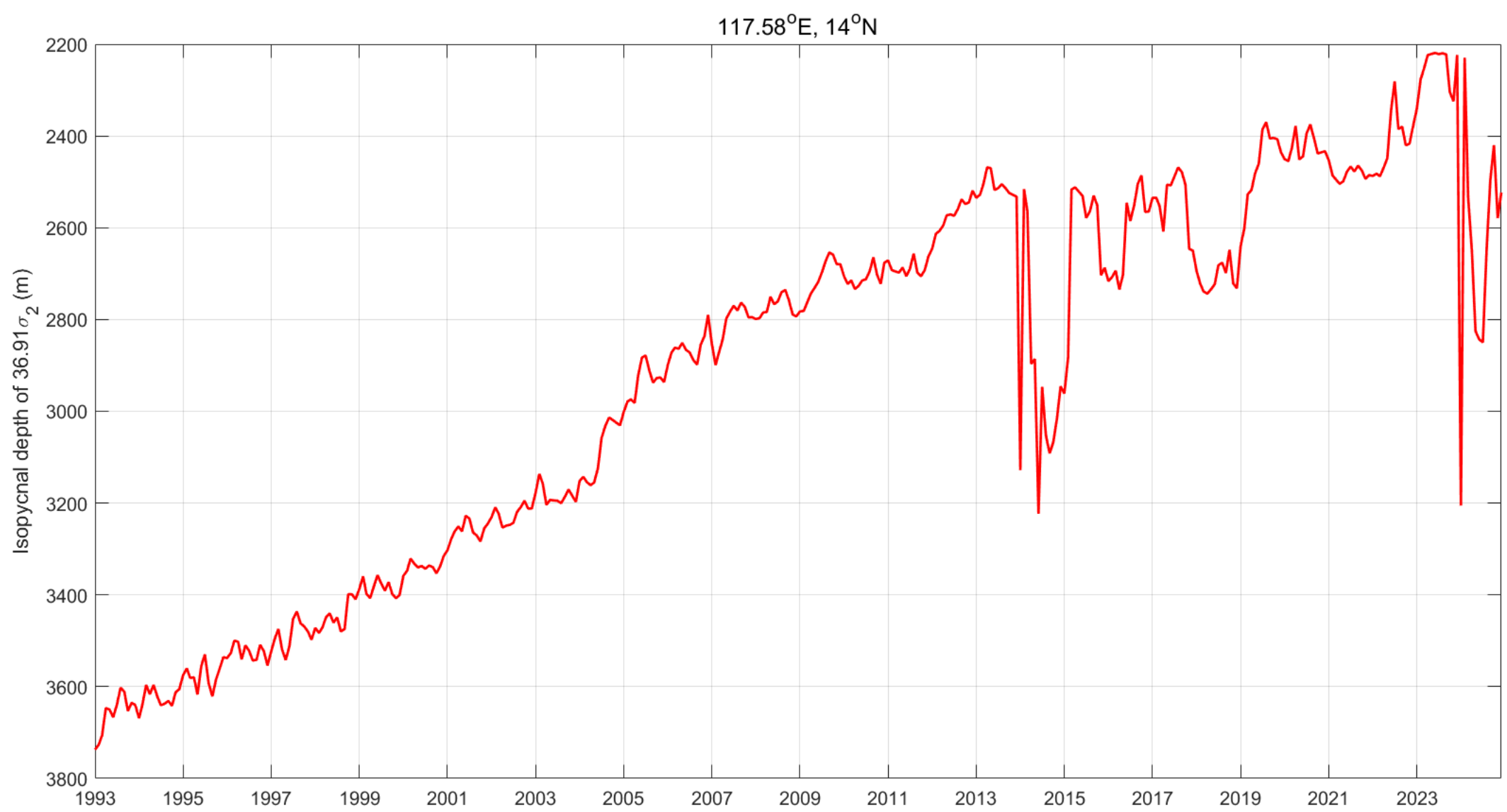

Figure S4. Time evolution of the depth of the 36.91 $\sigma_2$ isopycnal at 117.58°E, 14°N. Abrupt changes in isopycnal depth occur at several time points.